\documentclass{article}
\begin{document}

\begin{center}
\LARGE {The masses of the mesons and baryons.  \\ Part III.\quad The size of the 
particles}
\bigskip

\Large {E.L. Koschmieder} 
\medskip

\small{Center for Statistical Mechanics, The 
University of Texas at Austin \\ Austin, TX 78712,  USA \\ e-mail: 
koschmieder@mail.utexas.edu}
\smallskip

\large{May 28, 2000}
\end{center}



\bigskip
\noindent
\small{The size of the stable elementary particles is investigated with the 
standing wave model. The particle size follows from the magnitude of the 
radiation pressure. It is shown that the outward directed radiation 
pressure is balanced by the inward directed elastic force per unit area in the cubic 
nuclear lattice, provided that the sidelength of the lattice is 
\(10^{-13}\) cm, which agrees with the measured radius of the proton 
\begin{math} r = 0.8 \, \cdot \, 10^{-13}\end{math} cm, within the uncertainty of the 
parameters.}

\normalsize

\section{Introduction}
In a previous article [1] we have shown that it follows from the 
well-known decays and masses of the so-called stable elementary particles 
that the spectrum of the stable particles consists of a $\gamma$-branch 
and a neutrino branch, and that the masses of the $\gamma$-branch are 
integer multiples of the mass of the  \begin{math} \pi^0 \end{math}  meson, 
with an average deviation of 1.0059. In a following paper [2] we have 
explained the integer multiple rule with the eigenfrequencies of plane, 
standing, electromagnetic waves in a cubic nuclear lattice. We have found 
that the masses of the  $\gamma$ -branch, the  \begin{math} \pi^0,\,  
\eta\,,\,  
\eta^\prime\,,\,  \Lambda, \, \Sigma^0, \, \Xi^0, \, \Omega^-, \, 
\Lambda^+_c, \, \Sigma^0_c\,,  
\,\Xi^0_c\:\end{math}  and \begin{math} \, \Omega^0_c\, \end{math} 
particles, must be integer multiples of the mass of the $\pi^0$ meson. The 
ratios of the masses of these particles are, in the standing wave model, independent 
of the size of the lattice. However, when we 
determine the mass of the  \(\pi^0\)  meson in absolute terms, we need the 
number of the lattice points, which follows from the size of the lattice. 
In [2] we have used for the size of the lattice the known mean square 
radius of the proton, which is of order  \(10^{-13}\)  cm. We have, thereby, 
implied that the size of the \(\pi^0\) meson is the same as the size of 
the proton. In the following we will justify this assumption, showing that 
the size of the particles is limited to a particular value by the 
radiation pressure.

\section{Radiation pressure and Young's modulus}

Standing electromagnetic waves in a cubic lattice with free boundaries 
exert an outward directed force on the ends of the lattice, the radiation 
pressure, which is caused by the reflection of the waves at the lattice 
end. The lattice will break up, at the latest, when the radiation pressure is 
equal to the 
force per unit area by which one layer of the lattice attracts an adjacent 
layer. The radiation pressure is determined as follows: The force caused 
by the reflection of one wave from a boundary is given by the momentum 
change. It is $\Delta\,p = 2\,h\,/ \lambda$ and $\Delta\,t = 1 /\: 
\nu\;$, where  \emph{p}  stands as usual for momentum, \emph{h} for PlanckÕs quantum, $\lambda$ 
for the wavelength, \emph{t} for the time, and \emph{c} for the velocity of light. The force 
caused by the reflection of all permissible waves with the wavelengths 
$\lambda  = 2\,L \, /  \, n$, where \emph{L} is the sidelength of the lattice, and \emph{n} 
an integer number ranging from 1 to about 1000, is given by
\begin{equation}
\sum_n\,\frac{\Delta\,p_n}{\Delta\,t_n}=\sum_n\,\frac{2\,h}{\lambda_n}\:\nu_n=\sum_n
\,\frac{n\,E_n}{L} \; ,
\end{equation}
which, when applied to the entire lattice end, is 
\begin{equation}
F = m\,c^2/L\,,                  
\end{equation}
where \emph{m} is the mass of the particle. The force per unit area, or pressure \emph{P} is
\begin{equation}
P = m\,c^2/L^3 = \rho\,c^2 .
\end{equation}

The energy in the mass of the $\pi^0$ meson is 134.9 MeV = 2.16 $\cdot~10^{-4}\:$ erg.
Assuming that $L = 10^{-13}$ cm it follows that the radiation pressure acting on an end of the cubic lattice is
\begin{equation}
P = 2.16 \cdot 10^{35} \quad\mbox{dyn/cm$^2$} .			    
\end{equation}

The outward directed radiation pressure can at most be equal to the inward acting 
elastic force at the lattice end. The elastic force per unit area is characterized by Young's modulus \emph{Y}. 
We have previously determined the value of \/ Young's modulus of a cubic 
lattice held together by the weak nuclear force in [3], Eq.(21) therein. 
We found 
\begin{equation}
Y = 2.06 \cdot 10^{35} \quad\mbox{dyn/cm$^2$}.
\end{equation}

Within the uncertainties of the parameters, in particular of the breaking 
point of the lattice, the radiation pressure \emph{P} is 
balanced by the elastic force of the lattice, \emph{provided} that the sidelength 
of the lattice is $L = 10^{-13}$ cm. \emph{L} determined from either Eq.(4) or Eq.(5) 
differs by only 1.6\%. We will use in the following $L = 10^{-13}$ cm. 
If $|Y| = |P|$ 
we can write Eq.(3) as the well-known formula for the velocity of elastic 
waves in a rigid body
\begin{equation}
v = \sqrt{Y/\rho}.
\end{equation}
The velocity in the case of the nuclear lattice is equal to the velocity 
of light, which cannot be exceeded, nor can the velocity be smaller than 
\emph{c}, because the waves making up the particles of the $\gamma$-branch are 
electromagnetic.

   The considerations above apply to the lowest mode of the lattice 
oscillations, that is to the $\pi^0$ meson. The higher modes of the lattice 
oscillations correspond to the $\eta$ meson, $\Lambda$ baryon etc, whose masses are, 
within 3\%, integer multiples of the mass of the $\pi^0$ meson, as discussed in 
[1], in particular Table 1 therein. The outward directed radiation 
pressure of the higher modes must be equal to the value of the radiation 
pressure \emph{P} of the $\pi^0$ meson times an integer number, representing the ratio 
of the mass of the particular $\gamma$-branch particle divided by $m(\pi^0)$, because, 
as we have learned in [2], the higher modes of the standing waves in the 
nuclear lattice differ from the waves in the $\pi^0 $ meson only in the number 
of the waves,  not in the frequencies.  As the number of waves increases, 
so does the radiation pressure, provided that the sidelength of the 
lattice is the same for all particles of the $\gamma$-branch. But since the 
density $\rho  = m\, /  \, L^3$ of the particles with higher modes increases 
if \emph{L}~=~const, the velocity \emph{v} would, according to Eq.(6), 
decrease, if  Young's 
modulus \emph{Y} of the particles with the higher modes remains constant. 
However, since the velocity of the standing waves in a cubic nuclear 
lattice is the constant velocity of light, Eq.(6) requires that Young's 
modulus increases by the same amount as the density of the particles with 
the higher modes. Young's modulus of the particles does indeed increase 
this way, because as the number of waves increases, so does the number of 
strings which hold the lattice together, each wave corresponding to a 
string. In other words, in a higher mode the density $\rho$ increases as the 
number of waves increases, and the radiation pressure increases by the 
same amount for the same reason. Young's modulus increases likewise by the 
same amount, because the internal forces in the lattice are proportional 
to the number of waves in the lattice. The increase of \/ $\rho$, \emph{P} and 
\emph{Y} is the 
same, \emph{provided} that the sidelength of the lattice is independent of the 
order of the modes.

\section{Conclusion}

It is shown that the outward directed radiation pressure of standing, 
electromagnetic waves in a cubic nuclear lattice is balanced by the 
elastic force per unit area, provided that the sidelength of the lattice 
is equal to $10^{-13}$ cm, which means about equal to the size of the 
proton. The size of the particles is an automatic 
consequence of the standing wave model of the stable elementary particles 
of the $\gamma$-branch. Similar considerations apply for the particles of the 
$\nu$-branch which are held together by a neutrino lattice.
\bigskip

\noindent
\textbf{References}
\smallskip

\noindent
[1] E.L.Koschmieder, to appear in Proc. Roy. Belg. Acad. Sci.;\\xxx.lanl.gov/abs/hep-ph/0002179.
\smallskip

\noindent
[2] E.L.Koschmieder, to appear in Proc. Roy. Belg. Acad. Sci.;\\xxx.lanl.gov/abs/hep-lat/0002016.
\smallskip

\noindent
[3] E.L.Koschmieder, Nuovo Cim. \textbf{A 101},1017 (1989).

\end{document}